# Noise analysis in outdoor dynamic speckle measurement

MIKHAIL LEVCHENKO[1], ELENA STOYKOVA[1*], BRANIMIR IVANOV[1], LIAN NEDELCHEV[1], DIMANA NAZAROVA[1], KIHONG CHOI[2], JOONGKI PARK[2]

[1]*Institute of Optical Materials and Technologies, Bulgarian Academy of Sciences, Acad. Georgi Bonchev Str., Bl.109, 1113 Sofia, Bulgaria*
[2]*Electronics and Telecommunications Research Institute, 218 Gajeong-ro, Yuseong-gu, Daejeon, 34129, Republic of Korea*
*\*Corresponding author: estoykova@iomt.bas.bg*

**Dynamic speckle method is an effective tool for estimation of speed of processes. Speed distribution is encoded in a map built by statistical pointwise processing of time-correlated speckle patterns. For industrial inspection, the outdoor noisy measurement is required. The paper analyzes efficiency of the dynamic speckle method in the presence of environmental noise as phase fluctuations due to lack of vibration isolation and shot noise due to ambient light. Usage of normalized estimates for the case of non-uniform laser illumination is studied. Feasibility of the outdoor measurement has been proven by numerical simulations of noisy image capture and real experiments with test objects. Good agreement has been demonstrated both in simulation and experiment between the ground truth map and the maps extracted from noisy data. © 2022 Optica Publishing Group**

## 1. INTRODUCTION

Dynamic speckle method (DSM) enables visualization of speed distribution for processes that occur in a 3D object and lead to micro-changes of its relief [1,2]. The intensity-based DSM attracts with simplicity of implementation and shows increasing popularity. It requires laser illumination to form a speckle pattern on the object surface and a digital optical sensor to record a sequence of such patterns. Usually, this sequence is used for pointwise processing [3,4]. The output is a 2D map of a statistical parameter which, in general, is non-linearly related to the correlation radius of the temporal intensity fluctuations at different object points. The map is known as an activity map, because it characterizes the areas of faster or slower change on the object surface. The map as a 2D array consists of strongly fluctuating entries due to speckle nature of the raw data. Hence different algorithms have been proposed to improve the map contrast. Efficiency of the DSM has been also checked by solving various tasks in non-destructive testing as estimation of blood flow perfusion [5] and penetration of cosmetic ingredients [6], ear biometrics [7], measurement of bacterial response [8], observation of processes in plants [9,10], evaluation of seeds viability [11] and animal reproduction [12], food quality assessment [13,14], drying of paints [15] and polymer thin films [16], fire detection [17] to name a few.

For industrial or medical applications, the DSM-based non-destructive testing must be effective under outdoor conditions. This means data acquisition is performed under inevitably increased phase noise due to lack of vibration-isolated environment and in the presence of ambient light. The aim of this work is to study the impact of these two sources of environmental noise on the DSM output and to prove applicability of the method for outdoor measurements.

Noise analysis must take into account specifics of the DSM which provides information about i) spatial variation of the statistical parameter encoding the speed and ii) evolution of this variation in time. Accordingly, outdoor DSM implementation must also correctly indicate the relative speed distribution and its temporal evolution. The absolute values of the statistical parameter may differ in this case from its ground truth values observed for vibration isolation and ambient light excluded. In addition, outdoor acquisition of raw data in combination with frequently encountered case of non-uniform object illumination may lead to erroneous entries in the activity map. This is due to the need to apply normalized processing in order to compensate for the signal-dependent intensity variation. The outdoor acquisition should allow raw data compression which is mandatory when storage of large amount of data is required.

Noise analysis in the paper is done by processing data from numerical simulations and experiments with test objects. For numerical experiments, we take as a ground truth (GT) activity map the map obtained from data acquired with vibration isolation and no ambient light. From different sources of noise which characterize the capture process in CCD or CMOS cameras [18], we include in the simulation of GT raw data the shot noise related to detection of photons at the used laser wavelength and the noise produced by quantization at recording 8-bit encoded images. As a first step of the noise analysis, we modeled external noise as time-correlated phase added to the complex amplitude of the light field on the object surface and simulated the shot noise corresponding to the contribution of the ambient light. As a second step, we proved outdoor efficiency of the DSM by conducting drying experiments with different test objects. To characterize quality of the activity map as a DSM output, we use as a figure of merit the probability density function (PDF) of the estimate of the statistical parameter used for evaluation of activity.

The structure of the paper is as follows: in Sec.2 we describe the steps of simulation of noisy speckle images and analyze efficiency of three intensity-based pointwise algorithms suitable for uniform or non-uniform laser illumination in the presence of environmental noise. In Sec.3, we provide experimental verification of the option to perform the dynamic speckle measurement with vibration isolation turned off and presence of ambient light. Both Sec.2 and Sec.3 present results for grayscale speckle images.

## 2. NOISE ANALYSIS OF SYNTHETIC SPECKLE IMAGES
### A. Simulation of speckle images for outdoor acquisition

We consider intensity-based pointwise DSM when a 2D optical sensor, focused on a 3D object surface, captures a sequence of speckle patterns of size $N_x \times N_y$ pixels at a pixel interval, $\Delta$. Any two consecutive images in the sequence are separated by the time interval, $\Delta t$. From $N$ recorded images, temporal sequences of intensity values, $I_{kl,i} \equiv I(k\Delta, l\Delta, i\Delta t), i = 1..N$, are extracted for each pixel $(k,l) \equiv (k\Delta, l\Delta), k = 1..N_x, l = 1..N_y$. Note that the intensity values are integers in the interval from 0 to 255 for 8-bit encoded images. The extracted sequences are input data for different correlation-based algorithms, e.g. a modified structure function (MSF) [19]:

$$S_1(k,l,m) = \frac{1}{(N-m)} \sum_{i=1}^{N-m} |I_{kl,i} - I_{kl,i+m}| \qquad (1)$$

where the integer $m$ shows the time lag, $\tau = m\Delta t$, between the compared intensities. This MSF estimate, which is called Estimate 1 in the paper, provides maps with a good contrast both for symmetrical and asymmetrical speckle intensity distributions. The main drawback of Estimate 1 is its applicability only for uniform illumination of the object at equal reflectivity everywhere on its surface. If these conditions are not met, one should use a normalized estimate as e.g. Estimate 2 [20,21]:

$$S_2(i,k,m) = \frac{1}{(N-m)} \sum_{i}^{N-m} \frac{|I_{kl,i} - I_{kl,i+m}|}{(I_{kl,i} + I_{kl,i+m} + q)} \qquad (2)$$

The optional parameter $q$ is inserted in the algorithm to avoid division by zero in case of pixels with zero intensity.

The contour maps of estimates $S_1$ and $S_2$ visualize activity across the object within the averaging interval, $T = N\Delta t$. Spatial variation of activity can be described by the spatial distribution of the temporal correlation radius, $\tau_c(k,l)$, of intensity fluctuations on the object surface. The lower temporal correlation, the higher activity. Both estimates increase with activity and strongly fluctuate from point to point. The spread of the estimate fluctuations also increases with activity thus introducing signal dependent noise in the activity map. The condition for a good contrast map is to provide as narrow as possible PDF of the estimate at a given $\tau_c$. Important specifics of the pointwise DSM is that the averaging interval $T = N\Delta t$ can exceed or be less than $\tau_c(k,l)$ at a certain pixel or object point respectively.

Schematics of generation of synthetic speckle images in the presence of environmental noise is shown in Fig.1. The object is illuminated with an expanded laser beam with uniform or non-uniform distribution of intensity at wavelength, $\lambda$. We accepted that the scattering centers on the rough object surface move randomly in the two directions normal to the object surface as a result of some process ongoing in the object. We assumed that the amplitude and the phase of light scattered by a given center were mutually independent. They were also independent of the amplitudes and phases of the other scattering centers. We accepted also that no temporal change in reflectivity occurred across the object during acquisition of the raw data It was reasonable to describe the phase change related to change of the height of a scattering center due to the process in the object as normally distributed at each point [22]. The phase change, $\Delta\varphi_m^{kl}$, at point $(k\Delta, l\Delta)$ between the moments separated by a time lag $\tau = m\Delta t, m = 1,2...N_\tau < N$ makes intensity in the optical sensor to fluctuate with a normalized correlation function $\rho_{kl}(\tau = m\Delta t) = \exp(-\sigma^2\{\Delta\varphi_m^{kl}\})$ [22,23], where $\sigma^2\{\Delta\varphi_m^{kl}\}$ is the variance of the phase change. Without limiting the generality of choosing various models, we used for $\rho_{kl}(\tau = m\Delta t)$ the model $\rho_{kl}(\tau) = \exp[-\tau/\tau_c(k,l)]$. This model describes effectively many processes as e.g. paint drying. For two successive images in the recorded sequence, the following formula is obtained for the standard deviation of the phase change, $\sigma\{\Delta\varphi_{m=1}^{kl}\} = \sqrt{\Delta t/\tau_c(k,l)}$.

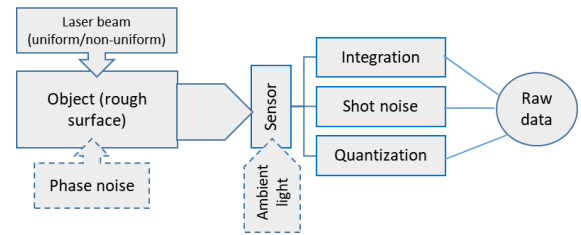

Fig. 1. Schematics of simulation of the raw data in dynamic speckle measurement in the presence of environmental noise.

Simulation of the scattered light started with generation of delta-correlated in space random phases on the object surface $\varphi(k\delta, l\delta, i\Delta t), k = 1..2N_x, l = 1..2N_y, i = 1..N$. The phase spatial distribution is generated at a spatial step $\delta = \Delta/2$ from a 2D array of phase values uniformly distributed from 0 to $2\pi$. The phase distribution at moment $i\Delta t$ for $i \geq 2$ is found at each point $(k\delta, l\delta)$ from the relation $\varphi(k\delta, l\delta, i\Delta t) = \varphi[k\delta, l\delta, (i-1)\Delta t] + \chi_{kl,i}\sqrt{\Delta t/\tau_c(k,l)}$ where $\chi_{k,li}$ is a random number with standard normal distribution with zero mean and variance equal to 1, $k = 1..2N_x, l = 1..2N_y, i = 1..N$. This number is separately generated for each combination of indices, $k, l, i$

The environmental phase noise is added at all moments $i\Delta t, i = 1..N$ starting from $i \geq 2$. We considered the case of independent noisy fluctuations at the object points. The noisy fluctuations had the same probability distribution at all points and the same temporal correlation radius, $\tau_{noise}$. Similarly to the phase change related to the observed process, we assumed that the standard deviation of the environmental phase fluctuations was given by $\sigma_{noise} = \alpha\sqrt{\Delta t/\tau_{noise}}$ where parameter α is less than 1. Parameter α was introduced to take into account that the spread of the noisy fluctuations is expected to be less than the phase change due to the observed process. The complex amplitude on the object surface was generated from the phase distributions $\varphi(k\delta, l\delta, i\Delta t)$ with added phase noise for intensity distribution $I_0(k\delta, l\delta)$ of the laser beam on the object surface at the instant, $i\Delta t$, as $U_S = \sqrt{I_0(k\delta, l\delta)}\exp\{-j[\varphi(k\delta, l\delta, i\Delta t) + \theta_{kli}\sigma_{noise}]\}$, where $j$ is the imaginery unit, $\theta_{kl,i}$ was a random number with standard normal distribution with zero mean and variance equal to 1. The random value $\theta_{kl,i}$ was also separately generated for each combination of indices, $k, l, i$. Note that at this stage of simulation, spatial intensity distribution $I_0(k\delta, l\delta)$ of the laser beam was not given by integer numbers.

Simulation proceeded further with generation of the complex amplitude of the light field on the sensor aperture $U_{cam} = FT^{-1}\{H \cdot FT\{U_S\}\}$ where $FT\{\cdot\}$ denotes Fourier transform and $H$ is a *circ* function in the Fourier domain with a cut-off frequency equal to $N_x\delta D/(2\lambda f)$, where $D$ and $f$ are the diameter and the focal distance of the camera objective and we assumed that $N_x = N_y$ [24]. By varying the parameters of the cut-off frequency, it is possible to generate speckle of different averages size and with low or high contrast.

Integration of speckle by the camera pixels was simulated by summation of values $|U_{cam}|^2$ in a window of size $2\times2$ pixels. No averaging of the speckle within the exposure interval was simulated. It was assumed that the exposure interval was much shorter than the interval $\Delta t$.

From the numerous sensor noise sources which accompany transformation of irradiance on the sensor entrance into number of photons, electrons, voltage and digital signal, we included in the simulation model the signal-dependent shot noise related to photons at the laser wavelength and the quantization noise at encoding the signal as 8-bit images. These two types of noise are the most substaltial at acquisition and inevitably exist in all raw data, even in those captured in darkened environment with vibration isolation. At outdoor capture, ambient light enters the sensor aperture and the photons at wavelengths within the curve of spectral sensitivity of the camera create additional shot noise. Therefore, detection was simulated at assumptions that i) the average number of signal photons arriving within the exposure interval with energy corresponding to the laser wavelength was $N_{ph}(l,k)$, ii) the maximum average number of signal photons within the exposure interval for the acquired images was $N_{max}$; iii) ambient light created equivalent average number of photons, $N_{al} = \eta N_{max}$, that was the same for all pixels. We assumed for the simulation that parameter $\eta$ was less than 1. The shot noise was modeled as random number with a Poisson distribution with average and variance equal to $N_{ph}(l,k) + N_{al}$. To determine $N_{ph}(l,k)$, the maximum intensity in the generated 2D speckle intensity distributions before the procedure of quantization was taken to correspond to $N_{max}$. The camera dynamic range was adjusted to cover a signal equivalent to $N_{max} + \sqrt{N_{max}}$. The simulated raw data were transformed to 8-bit encoded bitmap images for further processing.

**B. Noise analysis at constant activity**
The purpose of simulation by using the developed simplified model of different sources of environmental noise was to characterize the impact of these sources and to prove feasibility of the DSM in noisy environment.

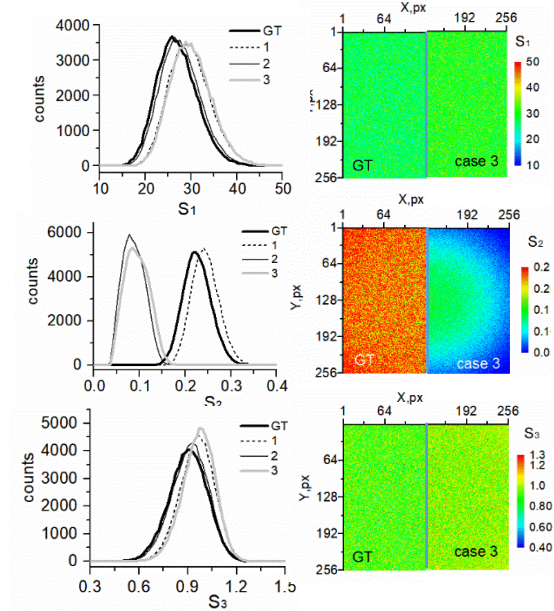

Fig. 2. Left: histograms of estimates $S_1$, $S_2$, $S_3$ built from the GT maps and noisy maps for uniform ($S_1$) and Gaussian ($S_2$, $S_3$) illumination: case 1 - $\alpha = 0.2; \eta = 0.0$, case 2 - $\alpha = 0.0; \eta = 0.2$, case 3 - $\alpha = 0.2; \eta = 0.2$; $\tau_{noise} = \Delta t$. Right: combined activity maps (GT map from 1 to 128 and a map for noise case 3 from 129 to 256) of estimates $S_1$ for uniform illumination and estimates $S_2$ and $S_3$ for Gaussian illumination.

As a first test object, we simulated a flat square surface with constant activity, i.e. $\tau_c = const$. The parameters of simulation were as follows: $N_x = N_y = 256$, $N = 200$, $\tau_c = 20\Delta t$, wavelength 632.8 nm, $\tau = 10\Delta t$, $N_{max} = 1000$. Simulation was done for uniform and Gaussian laser beam with $I_0(k\delta, l\delta) = \exp\{-\delta^2[(k-N_x)^2 + (l-N_y)^2]/\Omega^2\}$. The parameter $\Omega$ was equal to $300\Delta$. We simulated the data for the case of vibration isolation and no ambient light in order to build the GT map. For noise analysis, we modeled 3 cases: case 1 with $\alpha = 0.2; \eta = 0.0$ (vibration isolation off, no ambient light), case 2 with $\alpha = 0.0; \eta = 0.2$ (vibration isolation on, ambient light), and case 3 with $\alpha = 0.2; \eta = 0.2$ (vibration isolation off, ambient light). In all the three cases, we chose $\tau_{noise} = \Delta t$. Due to constant activity, it is possible to use all 256×256 entries in the activity maps to build histograms of the estimates $S_1$ and $S_2$ for evaluation of their PDF's at a given activity value.

The histograms of both estimates $S_1$ and $S_2$ for the GT map and the three simulated noise cases are shown in Fig.2 (left) for uniform illumination in order to characterize fully the impact of the phase noise and ambient light. We see rather different behavior of the normalized and non-normalized estimates in the presence of the environmental noise. As it should be expected, the non-normalized Estimate 1 ($S_1$) is weakly affected by the shot noise due to the ambient light. The standard deviation of the random number of the ambient light photons is about 14 photons which means that the contribution of the ambient light to the absolute difference $|I_{kl,i} - I_{kl,i+m}|$ is rather small for any two values of "i". The phase noise shifts the estimate distribution to slightly higher values without changing noticeably the shape of the histogram. The weak impact of the simulated noise on Estimate 1 is confirmed by the activity maps corresponding to $S_1$ in Fig.2 (right). The maps were obtained for uniform illumination. For better comparison, we plotted half of the GT map from $k = 1$ to $k = 128$ next to the half of the map for noise case 3 from $k = 129$ to $k = 256$. Both parts of the activity maps are plotted with the same scale.

For the normalized Estimate 2, the impact of the ambient light is very strong due to its strong influence on the denominator in Eq.(2). The impact is expressed as shift of the histogram to lower activity values as well as failure of normalization. This is clearly seen from comparison of the half of the GT map for Estimate 2 ($S_2$) to the half of the map for noise case 3 in Fig.2(right) when the maps are built for data acquired under Gaussian illumination. The GT map for this estimate is practically uniform despite the substantial decrease of intensity at the periphery of the images at Gaussian illumination. The map of Estimate 2 built from noisy raw data reveals the non-uniformity of illumination despite the normalization. The conclusion is that this normalized estimate fails in correct visualization of activity in the presence of ambient light. The phase noise at $\alpha = 0.2$ and $\tau_{noise} = \Delta t$ has much weaker impact expressed as shifting the histogram of the estimate to slightly higher values on the estimate axis.

The obtained result shows that a normalized estimate must be found with low sensitivity to the ambient light shot noise. We propose Estimate 3:

$$S_3(k,l,m) = \frac{1}{(N-m)} \sum_{i=1}^{N-m} \frac{1}{\sigma_{kl}} |I_{kl,i} - I_{kl,i+m}| \quad (3)$$

where the estimate of the standard deviation, $\sigma_{kl}$, is given by

$$\sigma_{kl}^2 = \frac{1}{N} \sum_{i=1}^{N} (I_{kl,i} - \bar{I}_{kl})^2, \quad \bar{I}_{kl} = \frac{1}{N} \sum_{i=1}^{N} I_{kl,i} \quad (4)$$

The shot noise is expected to have weak impact on the standard deviation, $\sigma_{kl}$, and this will improve efficiency of the algorithm. As is seen from Fig.2 (left), the shot noise caused by the ambient light leads to practically the same histogram as the GT histogram for uniform illumination. The phase noise affects the histogram of Estimate 3 to a larger extent. For Gaussian illumination, the GT map and the noisy map for case 3 in Fig.2 (right) show rather good similarity and non-uniformity of illumination is intractable.

**C. Map contrast for two levels of activity**

Applying the DSM in outdoor conditions is acceptable if only the output maps provide reliable information about the spatial variation of activity and evolution of activity in time. The areas of faster or slower change on the object surface must be visualized with good contrast.

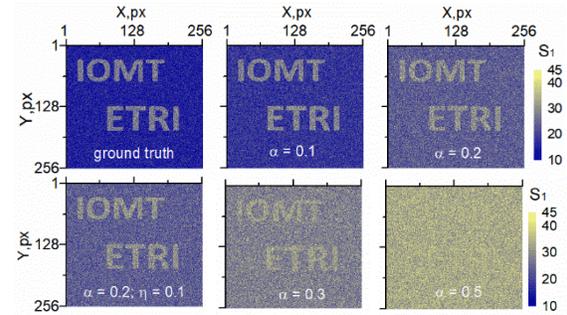

Fig. 3. Activity maps of Estimate 1 for an object with two activity levels (two logos and uniform background) at increasing phase noise for uniform illumination: $\tau_{noise} = \Delta t$, $\tau_{cl} = 10\Delta t$, $\tau_{cb} = 50\Delta t$, $N = 200$, $\tau = 10\Delta t$.

As a second task, we studied contrast of the activity maps obtained in noisy environment for a test object with two levels of activity and clear borders between both activity areas. We made simulation for the object we used before [25] that represents compact areas of a rapidly evolving process within background with slow variation of intensity. The high activity areas were formed by the letters of the logos "IOMT" and "ETRI" of both research institutions which participated in the current study. The logos occupy 29490 pixels whereas 232654 pixels contain background intensity in the initial $2N_x \times 2N_y$ array of delta-correlated phases. Therefore, this object, as we have mentioned in [25], is appropriate for testing the DSM detection of relatively small activity areas buried in a background.

We simulated both uniform and non-uniform Gaussian illumination with the following simulation parameters: $N_x = N_y = 256$, $N = 200$, wavelength 632.8 nm, time lag $\tau = 10\Delta t$, $N_{max} = 1000$, temporal correlation radii for the logos and the background equal to $\tau_{cl} = 10\Delta t$ and $\tau_{cb} = 50\Delta t$ respectively, $\Omega$ = 300δ, and increasing $\tau_{noise}$.

The activity maps built for Estimate 1 at increasing level of the phase noise given by the parameter α at $\tau_{noise} = \Delta t$ are shown in Fig.3. The coordinates of a point in the map along the X and Y axes are given in pixels. For better visualization, since the maps are of the same dimensions, only one X axis is used for the maps in a single column and one Y axis for the maps in a single row. For this estimate, simulation was made for uniform illumination. All maps are plotted with the same scale for better comparison. The good contrast of the GT map is gradually worsening when the phase noise standard deviation goes up. At α = 0.3, the logos are still distinguishable although the map is barely acceptable. At α = 0.5, no logos are seen. The maps can be smoothed in order to improve the contrast at slight loss of spatial resolution. With maps smoothed or not smoothed, the DSM is completely applicable at $\alpha \leq 0.2$. This quantitative result is a base for the general conclusion that the DSM is feasible for outdoor conditions at rather considerable phase noise levels. Really, at α = 0.2, the ratio between the standard deviations of the phase change caused by the observed process and the phase change due to the noise is equal to $5(\tau_{noise}/\tau_{cl,b})^{1/2}$ for the area of the logos and the background respectively. At $\tau_{noise} = \Delta t$, this ratio is equal approximately to 1.6 for the area of the logos and to 0.71 for the background. Ambient light is not an obstacle when normalization of the algorithm is not required as it is seen in Fig.3.

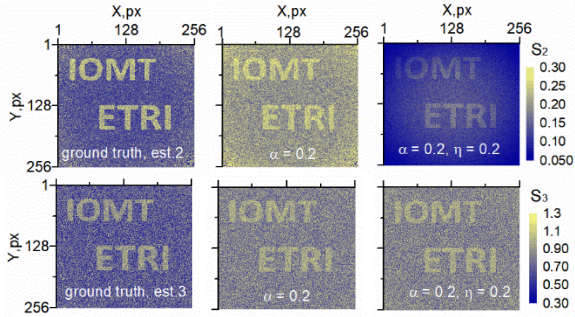

Fig. 4. Activity maps of Estimate 2 (top) and Estimate 3 (bottom) for an object with two activity levels (two logos and background) for Gaussian illumination at phase noise and ambient light shot noise: $N$ = 200, $\tau_{noise} = \Delta t$, $\tau_{cl} = 10\Delta t$, $\tau_{cb} = 50\Delta t$, $\tau$ = 10 $\Delta t$.

The activity maps built for Estimate 2 and Estimate 3 for Gaussian illumination are presented in Fig.4. Again the maps are plotted with the same scale. The GT maps are compared to the map extracted from the data with only phase noise and a map obtained from data acquired with phase noise and ambient light. The failure of Estimate 2 to reconstruct properly the activity at Gaussian illumination and ambient light is clearly seen: the map of Estimate 2 at $\alpha = 0.2; \eta = 0.2$ shows substantial shift to lower activity values and fall of activity from the center to the borders of the map. Estimate 3 reconstructs correctly activity variation but the contrast of the maps built from noisy data is comparatively low.

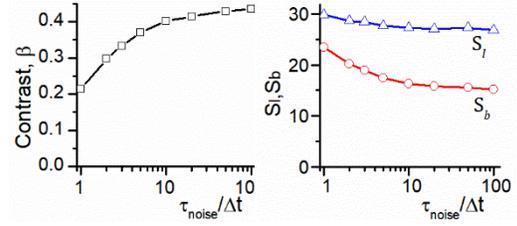

Fig. 5. Contrast parameter, β, for activity maps of Estimate 1 (left) and mean activity values in the high activity area ($S_l$) and low activity area ($S_b$) as a function of the correlation radius of the phase noise: $\tau_{cl} = 10\Delta t$, $\tau_{cb} = 50\Delta t$, $N$ = 200, $\tau$ = 10$\Delta t$, uniform illumination.

We introduced a sensitivity or contrast parameter, $\beta = (S_l - S_b)/S_l$, to characterize quality of the activity map, where $S_l, S_b$ are the mean values of the used estimate in the areas of the logos and the background, respectively. Due to the pointwise processing, these mean values were easily determined from the maps by using binary masks which extract only the area under the logos or the background. We studied the contrast parameter in the general case of environmental phase noise with some correlation radius, $\tau_{noise}$, in the time domain. The results obtained for Estimate 1 are shown in Fig.5. Each point in the plots corresponds to separate simulation of 200 images at a given value of $\tau_{noise}$. All sequences of images were generated at $\alpha = 0.2; \eta = 0.0$ as only this type of noise is substantial for the non-normalized estimate. As it should be expected, the contrast of the activity map is improving with $\tau_{noise}$ increasing, and β approaches the value corresponding the GT map.

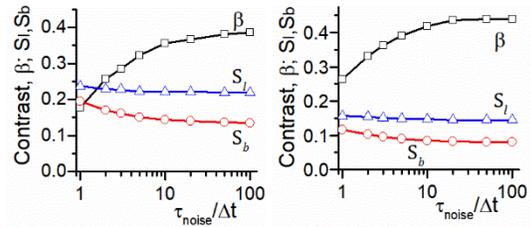

Fig. 6. Contrast parameter, $β$, and mean activity values in the high activity area ($S_l$) and low activity area ($S_b$.) for activity maps of Estimate 2 at $\alpha = 0.2; \eta = 0.0$ (left) and $\alpha = 0.2; \eta = 0.1$ (right) as a function of the correlation radius of the phase noise: $\tau_{cl} = 10\Delta t$, $\tau_{cb} = 50\Delta t$, $N$ = 200, $\tau$ = 10 $\Delta t$, Gaussian illumination.

The results of the sensitivity study for Estimate 2 at Gaussian illumination are presented in Fig.6 for the case of only phase noise $\alpha = 0.2; \eta = 0.0$ (left) and the case of phase noise and ambient light $\alpha = 0.2; \eta = 0.1$ (right). The same increase of the contrast is observed with $\tau_{noise}$ going up. The ambient light substantially decreases the mean values but the sensitivity parameter remains high.

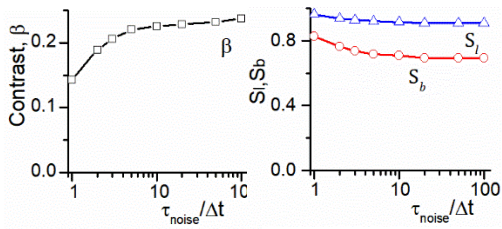

Fig. 7. Contrast parameter, *β*, and mean activity values in the high activity area ($S_l$) and low activity area ($S_b$) for activity maps of Estimate 3 at $\alpha = 0.2; \eta = 0.0$ as a function of the correlation radius of the phase noise: $\tau_{cl} = 10\Delta t$, $\tau_{cb} = 50\Delta t$, $N = 200$, $\tau = 10\Delta t$, Gaussian illumination.

The contrast parameter of Estimates 1 and 2 is practically two times higher than for Estimate 3 (Fig.7). The latter shows good stability in the presence of ambient light shot noise but worse contrast of the map in comparison to the other two estimates. This means that Estimate 2 may be a preferable choice in many cases when the non-uniformity of the intensity distribution is not so strongly expressed and contribution of the ambient light is small. The performed simulation analysis indicates that the faster phase noise fluctuations cause stronger impact on quality of the activity map.

## 3. EXPERIMENTAL NOISE ANALYSIS

The analysis in the previous section was based on a simplified model of the main noise sources which contribute into the raw data during the capture process under outdoor conditions. In this section, the issue of reliability of the outdoor DSM is approached by conducting drying experiments with several test objects printed by a 3D printer. The objects are schematically depicted in Fig.8 and described below. The experimental setup for capturing speckle patterns is shown in Fig.9. The object under study was placed on a vibration-isolated table. An expanded beam of linearly polarized laser light at wavelength 632.8 nm illuminated the sample. The laser power density, evaluated by powermeter PM100D from THORLABS, was equal to 0.67 mW/cm² on the object surface. A camera captured scattered coherent light with intensity fluctuating in time due to activity in the sample. We used acA4096-30um Basler camera with 4096 px x 2168 px resolution with pixel pitch 3.45 μm and exposure time 20 ms. Speckle images were captured at time interval Δ*t* = 2 s. A laptop computer was used in order to set recording parameters and to store the raw speckle images.

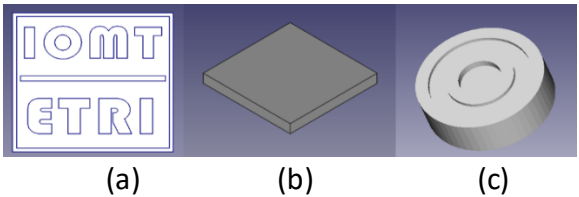

Fig. 8. Schematic representation of test objects.

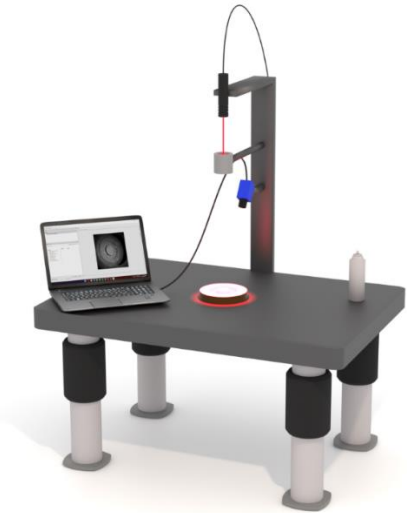

Fig. 9. Experimental setup for dynamic speckle measurement.

The first test object in Fig. 8(a) had the following dimensions: height 10 mm, width 40 mm, length 40 mm and depth of the letters 3 mm. The middle line size was as follows: length 36 mm, width 1.5 mm and depth 1 mm. The hollow regions formed the logos "IOMT" and "ETRI". The hollow regions and the flat object surface were covered with Acrylic paint Tamiya Mini X-2 White to create a smooth layer. This paint was suitable for either brush or spray application. In this and in the other experiments we used a brush. We conducted two experiments with and without ambient light and vibration isolation of the table turned on. The laptop computer was removed from the table during the experiments with vibration isolation on. We recorded two series of 1200 images with the same object which was covered anew with paint two times. The time series were divided into 6 sets of 200 images. The activity maps obtained for these sets with and without ambient light are shown in Fig.10 and Fig.11. The numbers along the horizontal and vertical axes of the maps show coordinates in pixels. Due to the slight non-uniformity of illumination, we used Estimate 2 for processing at a time lag equal to 5 Δ*t*. The maps in both cases exhibit similar behavior. They change from a map with more or less uniformly distributed high activity corresponding to active drying of the paint across the whole object surface in the beginning to maps with higher activity only within the regions of letters which still contain paint at the end of the experiment. The observed non-uniformity of activity in the background area is most probably due to the not very uniform deposition of paint using brush. The only difference between the maps is in the average value of Estimate 2. As it should be expected, this value is lower in the case of ambient light due to increased shot noise. This experiment confirms that the drying process can be reliably monitored in the presence of ambient light.

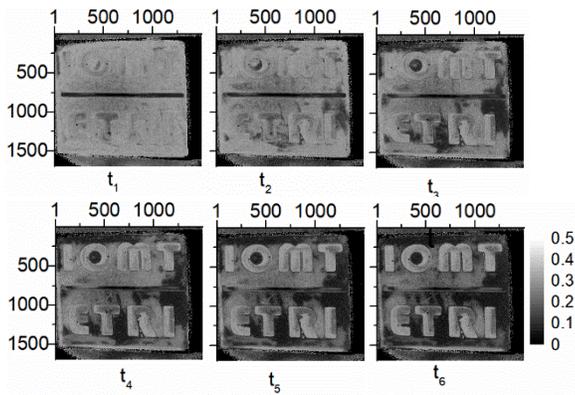

Fig. 10. Activity maps of Estimate 2 for a 3D object covered with paint with a flat surface and hollow regions in a darkened room and vibration isolation turned on; $t_1 = 0$ s, $t_2 = 6$ min 40 s, $t_3 = 13$ min 20 s, $t_4 = 20$ min, $t_5 = 26$ min 40 s, $t_6 = 33$ min 20 s, $N = 200$, $\tau = 5\Delta t$. The coordinates along the horizontal and vertical axes of the maps are in pixels.

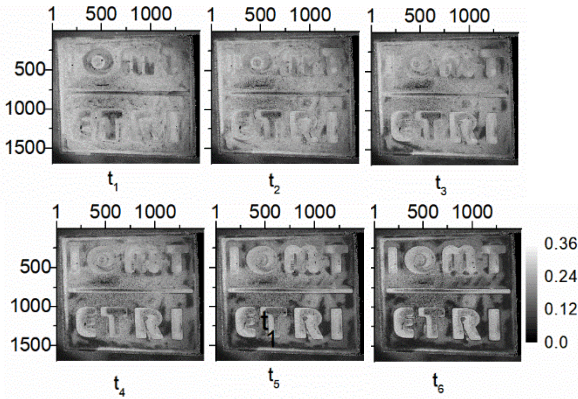

Fig. 11. Activity maps of Estimate 2 for a 3D object covered with paint with a flat surface and hollow regions in the presence of ambient light and vibration isolation turned on; $t_1 = 0$ s, $t_2 = 6$ min 40 s, $t_3 = 13$ min 20 s, $t_4 = 20$ min, $t_5 = 26$ min 40 s, $t_6 = 33$ min 20 s, $N = 200$, $\tau = 5\Delta t$. The coordinates along the horizontal and vertical axes of the maps are in pixels.

The next experiment was done with the flat plate shown in Fig. 8(b). We used a square plate of size 20 mm by 20mm and height 2 mm. The plate was covered anew with the same paint three times. Three measurements were conducted: in a darkened room with vibration isolation of the table turned on and in a room with ambient light and vibration isolation on and off. For each experiment 1000 images were acquired. The drying process of uniformly distributed layer of paint provides at least theoretically constant activity on the plate. We characterize this process by building histograms of Estimate 2 as a function of time. For the purpose, we chose $N = 200$ images to calculate the estimate and shifted this window of 200 images by one image at a time starting from the first image and going to the 800-th image. Thus we obtained 800 histograms of Estimate 2 that show decrease of activity due to drying. The results are shown in Fig.12. We see that, in the three studied cases, the decrease of activity with time follows the same dependence. The only difference is that the histograms for the measurements in the presence of ambient light are shifted to the lower activity values. This shift is clearly seen from the histograms corresponding to the beginning of the experiments that are also plotted in Fig.12. The histograms show the same behavior as the histograms for Estimate 2 in Fig.2 which depicts the theoretical predictions.

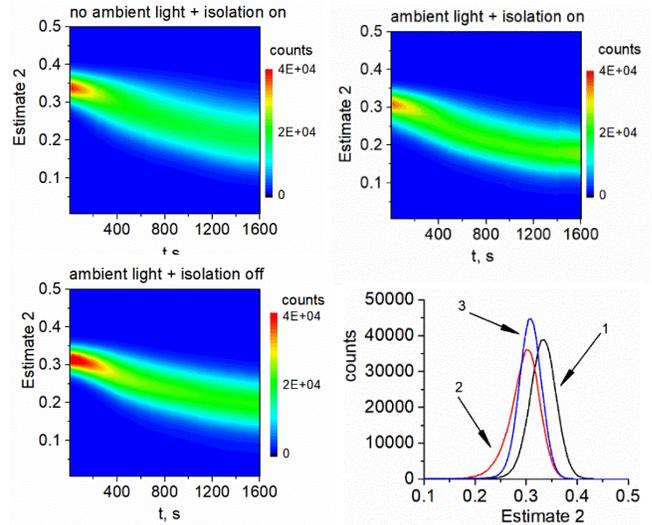

Fig. 12. Temporal variation of histograms and initial histograms of estimate 2 for a flat object covered with paint for measurement in a dark room and vibration isolation on (1), ambient light and isolation on (2) and ambient light and isolation off (3); $N = 200$, $\tau = 10\,\Delta t$.

The third experiment was conducted for the circular test object in Fig.8(c). The radii of the outer ring were 15 mm and 12 mm, the radii of the inner ring were 8 mm and 5 mm respectively. The height of the object was 10 mm and the depth of the hollow regions was 2 mm. To simulate working conditions in industrial environment, we applied the generator of vibrations, installed in smartphone Xiaomi Mi 9T. The smartphone was placed near the leg of the table with the experimental setup that was closest to the controlled sample. The generator was turned on during the entire experiment. We acquired two series of 900 images for i) no ambient light and vibration isolation turned on as well as for ii) ambient light, vibration isolation turned off and vibration generator turned on. The circular object was covered with paint anew two times using a brush. We used Estimate 3, and the activity maps corresponding to 6 different moments are shown in Fig.13 and Fig.14. Each map was computed from $N = 50$ images at a time lag equal to $4\,\Delta t$. The maps in Figs.13 and 14 clearly visualize consecutive stages of the drying process. The hollow regions are practically obscured on the first map in both figures, and they are well outlined at the end of the observed drying. The maps obtained with the vibration generator turned on are of very good contrast. These maps strongly resemble the GT maps in Fig.13, as is expected for Estimate 3.

The last experiment we conducted was monitoring of a polymer solution drying. We prepared methanol solution of the azopolymer poly [1- [4- (3-carboxy-4-hydroxyphe-nylazo) benzene-sulfonamido]-1,2-ethanediyl, sodium salt] -

PAZO which is widely used for recording of polarization holographic gratings. Drying of solutions of this polymer was monitored by us before [16]. In the current experiment, we captured the speckle images with and without the environmental noise. We dissolved 12 mg of PAZO in 250 μl methanol and deposited this solution on a glass plate using a micro-pipette. The aim was to have two identical plates with a polymer solution. The size of the plates was 24 mm × 36.5 mm. The glass thickness was 1.5 mm. The results of the experiment are shown in Fig.15. In the same way as in Fig. 12, we built the histograms of Estimate 2 for 800 activity maps computed from N = 200 images with successively increasing by 1 initial image. We also presented decreasing of the maximal number of the histograms in time. Note that the values of Estimate 2 are lower for capture with ambient light and phase noise than the GT values. At the same time, the maximal count for histograms from the noisy data is larger. Although two different plates have been used, the results obtained from noisy data are in a good agreement with the ground truth.

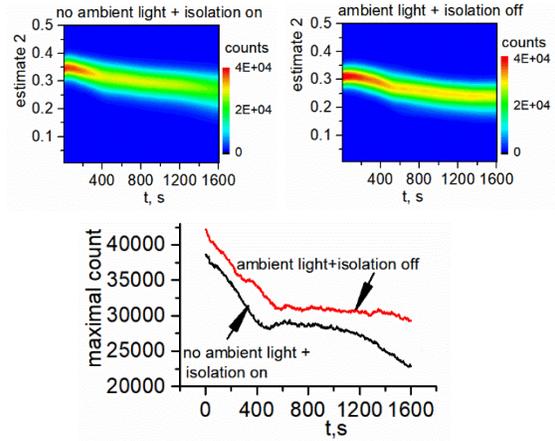

Fig. 15. Temporal variation of histograms of Estimate 2 (top) for a glass plate covered with polymer solution for measurement in a darkened room and vibration isolation (top left) and in presence of ambient light at vibration isolation off (top right); maximal count in the histograms (bottom) as a function of time; $N = 200$, $\tau = 10\,\Delta t$.

## 4. CONCLUSIONS

In summary, on the base of simulation and experiments, we have proved that the outdoor acquisition provides reliable data for intensity-based pointwise dynamic speckle analysis. These data are sequences of images of speckle patterns formed on the tested object surface under laser illumination. The output of the data processing is a 2D map of statistical estimate which is usually computed from dozens of speckle images. Despite the strong fluctuations of the map entries, it visualizes the spatial distribution of the degree of temporal correlation in the input data. A set of maps built at consecutive instants enables determination of the dynamics of a process. This makes the DSM an attractive approach in optical metrology, especially in view of its simple experimental realization. Analysis of the DSM stability to noise sources is an integral part of this realization and answers the question how effective the DSM is under outdoor conditions. Obviously, it is applicable to a certain level of noise and it is crucial to determine this level.

Noise analysis made in the paper studied the impact of i) the environmental phase noise due to lack of vibration isolation and ii) the increased shot noise due to ambient light. The GT map in this study was the map built from digitally acquired data in a darkened room with a setup positioned on vibration-isolated table. From the noise sources affecting the GT map, we included in analysis the shot noise from the laser photons and quantization noise. The validation of the outdoor measurement took into account the two requirements of the DSM implementation. The first is the mainly qualitative character of the DSM output. The activity map simply reproduces the areas of fast and slow change, and whenever these areas are distinguishable, the map is qualified as useful. The second criterion is correct description of the map evolution over time. We proved in the paper the fulfillment of both criteria in the presence of environmental noise. We determined by simulation the standard deviation of the phase noise above which the information in the map was destroyed.

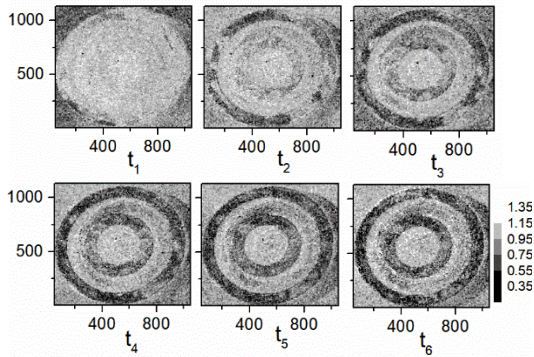

Fig. 13. Activity maps of Estimate 3 for a circular 3D object covered with paint and consisting of a flat surface and hollow regions for a darkened room at vibration isolation; $t_1 = 0$ s, $t_2 = 5$ min, $t_3 = 10$ min, $t_4 = 15$ min, $t_5 = 20$ min, $t_6 = 25$ min, $N = 50$, $\tau = 4\,\Delta t$.

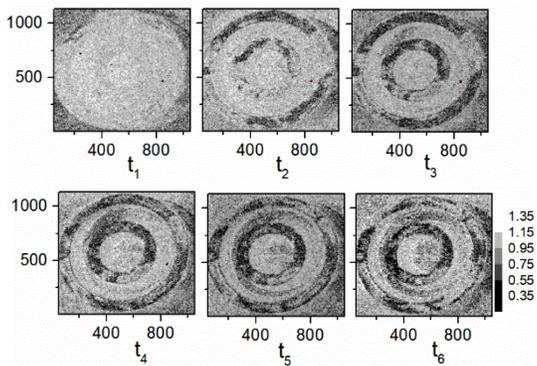

Fig. 14. Activity maps of Estimate 3 for a circular 3D object with paint and consisting of a flat surface and hollow regions in the case of ambient light at vibration generator turned on; $t_1 = 0$ s, $t_2 = 5$ min, $t_3 = 10$ min, $t_4 = 15$ min, $t_5 = 20$ min, $t_6 = 25$ min, $N = 50$, $\tau = 4\Delta t$.

.

We showed that the phase noise impact was decreasing if this noise was time-correlated. We found the normalized estimate for processing images acquired under non-uniform illumination with low sensitivity to ambient light. In a set of drying experiments with a phase noise and ambient light, we obtained the activity maps evolution similar to that observed for the GT maps.

**Funding** Institute of Information and Communications Technology Planning and Evaluation (IITP) grant funded by the Korea Government (MSIT) (2019-0-00001, Development of Holo-TV Core Technologies for Hologram Media Services)

**Acknowledgments** M. Levchenko thanks 2020 Plenoptic Imaging project for supporting his PhD training. This project has received funding from the European Union's Horizon 2020 research and innovation programme under the Marie Skłodowska-Curie grant agreement No 956770. E. Stoykova thanks European Regional Development Fund within the Operational Programme "Science and Education for Smart Growth 2014–2020" under the Project CoE "National centre of Mechatronics and Clean Technologies" BG05M2OP001-1.001-0008.

**Disclosures** The authors declare no conflicts of interest related to this article.

**Data Availability** Data underlying the results presented in this paper are not publicly available at this time but may be obtained from the authors upon reasonable request.

## References

1. J. Goodman, [Speckle Phenomena in Optics: Theory and Applications] Roberts and Company Publishers, (2007).
2. H. J. Rabal, and R. A. Braga, Jr., eds., [Dynamic Laser Speckle and Applications] CRC Press, (2009).
3. A.V. Saúde, F. S. de Menezes, P. L. Freitas, G. F. Rabelo, and R. A. Braga, Jr., "Alternative measures for biospeckle image analysis," J. Opt. Soc. Am. A **29**(8), 1648–1658 (2012).
4. E. Stoykova, B. Ivanov, and T. Nikova, "Correlation-based pointwise processing of dynamic speckle patterns," Opt. Lett. **39**(1), 115-118 (2014).
5. V. Rajan, B. Varghese, T. G. van Leeuwen, and W. Steenbergen, "Speckles in laser Doppler perfusion imaging," Opt. Lett. 31(4), 468–470 (2006).
6. E. Stoykova, B. Blagoeva, D. Nazarova, L. Nedelchev, T. Nikova, N. Berberova, Y. Kim, and H. Kang, "Evaluation of temporal scales of migration of cosmetic ingredients into the human skin by two-dimensional dynamic speckle analysis," Opt. Quant.Electron. 50, 191-201 (2018).
7. A. Chatterjee, P. Singh, V. Bhatia, and S. Prakash, "Ear biometrics recognition using laser biospeckled fringe projection profilometry," Opt. Las. Technol. **112**, 368-378 (2019).
8. B. Mandracchia, J. Palpacuer, F. Nazzaro, V. Bianco, R. Rega, and P. Ferraro, "Biospeckle decorrelation quantifies the performance of alginate-encapsulated probiotic bacteria", IEEE Journal of Selected Topics in Quantum Electronics **25**(1), 1-6 (2018).
9. R. A. Braga, L. Dupuy, M. Pasqual, and R. R. Cardoso, "Live biospeckle laser imaging of root tissues," Eur. Biophys. J. **38**(5), 679–686 (2009).
10. B. Ivanov, E. Stoykova, N. Berberova, T.Nikova, E. Krumov, and N. Malinowski, "Dynamic speckle technique as a leaf contamination sensor", Bulg. Chem. Commun. **45**(3), 149 – 153 (2013)
11. R. Braga, G. Rabelo, L. Granato, E. Santos, J. Machado, R. Arizaga, H. Rabal, and M. Trivi, "Detection of fungi in beans by the laser biospeckle technique," Biosystems Eng. **91**(4), 465-469 (2005).
12. R. Macedo, J. Barreto Filho, R. Braga, Jr, and G. Rabelo, "Sperm motility decreasing and semen fertility in the bull evaluated by biospeckle," Reprod. Fertil. Dev. **22**, 170–171 (2009).
13. C. Mulone, N. Budini, F. Vincitorio, C. Freyre, A. López Díaz, and A. Ramil Rego,"Analysis of strawberry ripening by dynamic speckle measurements," Proc. SPIE **8785**, 87851X (2013).
14. T. Lyubenova, E. Stoykova, E. Nacheva, B. Ivanov, I. Panchev, and V. Sainov, "Monitoring of bread cooling by statistical analysis of laser speckle patterns," Proc. SPIE **8770**, 87700S (2013).
15. A. Pérez, R. González-Peña, R. Braga Jr., Á .Perles, E . Pérez–Marín, F. García-Diego, " A Portable dynamic laser speckle system for sensing long-term changes caused by treatments in painting conservation,".Sensors **18** (1), 190 – 2-3 (2018).
16. R. Harizanova, V. Gaydarov, G. Zamfirova, E. Stoykova, D. Nazarova, B. Blagoeva, and L. Nedelchev, "Probing of the mechanical properties and monitoring of the drying process of azopolymer thin films for optical recording," Thin Solid Films **687**, 137441 (2019)
17. C. Christensen, Y. Zainchkovskyy, S. Barrera-Figueroa, A. Torras-Rosell, G. Marinelli, K. Sommerlund-Thorsen, J. Kleven, K. Kleven, E. Voll, J. Petersen, and M. Lassen, "Simple and robust speckle detection method for fire and heat detection in harsh environments," Appl. Opt. **58**, 7760-7765 (2019).
18. M. Konnik, and J. Welsh, "High level numerical simulations of noise in CCD and CMOS photosensors: review and tutorial." arXiv preprint arXiv:1412.4031 (2014).
19. E. Stoykova, D. Nazarova, N. Berberova, and A. Gotchev, "Performance of intensity-based non-normalized pointwise algorithms in dynamic speckle analysis," Opt. Express 23(19), 25128-25142 (2015)
20. H. Fujii, K. Nohira, Y. Yamamoto, H. Ikawa, and T. Ohura, "Evaluation of blood flow by laser speckle imaging sensing Part I," Appl. Opt. 26(24), 5321–5325 (1987).
21. E. Stoykova, "Preprocessing of raw data for quality enhancement of the pointwise dynamic speckle analysis", Proc. SPIE 10834, Speckle 2018: VII International Conference on Speckle Metrology 108341O (7 September 2018).
22. T. Fricke-Begemann, G.Gülker, K. D. Hinsch, and K. Wolff, "Corrosion monitoring with speckle correlation," Appl. Opt. 38, 5948-5955 (1999).
23. A. Federico, G. H. Kaufmann, G. E. Galizzi, H. Rabal, M. Trivi, and R. Arizaga, "Simulation of dynamic speckle sequences and its application to the analysis of transient processes," Opt. Commun. 260(2), 493–499 (2006).
24. E. Equis and P. Jacquot, "Simulation of speckle complex amplitude: advocating the linear model," Proc. SPIE 6341, 634138 (2006)
25. E. Stoykova, B. Blagoeva, N. Berberova-Buhova, M. Levchenko, D. Nazarova, L. Nedelchev, and Joongki Park, "Intensity-based dynamic speckle method using JPEG and JPEG2000 compression," Appl. Opt. **61**, B287-B296 (2022)